\documentclass[acmsmall,screen]{acmart}
\AtBeginDocument{%
  }

\setcopyright{acmlicensed}
\copyrightyear{2025}
\acmYear{2025}
\acmDOI{XXXXXXX.XXXXXXX}
\acmConference[FSE '26]{The ACM International Conference on the Foundations of Software Engineering}{July 05--09, 2026}{Montreal, QC}
\acmISBN{978-1-4503-XXXX-X/2025/09}

\usepackage{amsmath,amsfonts} 
\usepackage{graphicx}
\usepackage{textcomp}
\usepackage{xcolor}
\usepackage[linesnumbered,ruled,vlined,algo2e]{algorithm2e}
\usepackage{float}
\usepackage{adjustbox}
\usepackage{currfile}
\usepackage{listings}
\usepackage{multirow} % 用于合并行
\usepackage{makecell}
\usepackage{booktabs}
\usepackage{array}
\usepackage{stfloats} 
\usepackage{csquotes} % 允许使用直双引号自动转换 mk 20250420
\usepackage{epigraph} 
\usepackage{stfloats} 
\usepackage{tabularx} 
\usepackage{pifont}
\usepackage{cancel}
\usepackage{threeparttable} % 添加表格注释
\usepackage{xltabular}
\usepackage{booktabs}   % 提供更美观的表格线
\usepackage{ragged2e}   % 提供更好的文本对齐控制
\usepackage{rotating}
\usepackage{geometry} % 用于调整页面边距

\MakeOuterQuote{"} % 允许使用直双引号自动转换 mk 20250420
\newcommand{\sysname}[0]{\textsc{Matus}}

\definecolor{bluetagcolor}{RGB}{55,101,157}
 
\iftrue
\newcommand{\semihide}[1]{\textcolor{gray}{#1}}
\else 
\newcommand{\semihide}[1]{}
\fi

\begin{document}


% \title{
% Feature Slice Matching for Precise Bug Detection
% }
\title{Feature Slice Matching for Precise Bug Detection}


\author{Ke Ma}
\email{make@ruc.edu.cn}
\orcid{0009-0005-5732-0122}
\affiliation{%
 \institution{Renmin University of China}
 \city{Beijing}
 \country{China}
}

\author{Jianjun Huang}
\email{hjj@ruc.edu.cn}
\orcid{0000-0003-4403-0060}
\affiliation{%
 \institution{Renmin University of China}
 \city{Beijing}
 \country{China}
}

\author{Wei You}
\email{youwei@ruc.edu.cn}
\orcid{0000-0003-1009-6627}
\affiliation{%
 \institution{Renmin University of China}
 \city{Beijing}
 \country{China}
}

\author{Bin Liang}
\authornote{Corresponding author}
\email{liangb@ruc.edu.cn}
\orcid{0000-0002-4818-7164}
\affiliation{%
 \institution{Renmin University of China}
 \city{Beijing}
 \country{China}
}

\author{Jingzheng Wu}
\email{jingzheng08@iscas.ac.cn}
\orcid{0000-0001-5561-9829}
\affiliation{%
 \institution{Institute of Software, Chinese Academy of Sciences}
 \city{Beijing}
 \country{China}
}

\author{Yanjun Wu}
\email{yanjun@iscas.ac.cn}
\orcid{0000-0002-1823-045}
\affiliation{%
 \institution{Institute of Software, Chinese Academy of Sciences}
 \city{Beijing}
 \country{China}
}

\author{Yuanjun Gong}
\email{yuanjun.gong@unitn.it}
\orcid{0000-0002-4661-904X}
\affiliation{%
 \institution{University of Trento}
 \city{Trento}
 \country{Italy}
}

\renewcommand{\shortauthors}{Ma et al.}

\begin{abstract}
Measuring the function similarity to detect bugs is effective, but the statements unrelated to the bugs can impede the performance due to the noise interference. 
Suppressing the noise interference in existing works does not manage the tough job, i.e., eliminating the noise in the targets. 
In this paper, we propose \sysname{} to mitigate the target noise for precise bug detection based on similarity measurement. 
Feature slices are extracted from both the buggy query and the targets to represent the semantic feature of (potential) bug logics. 
In particular, \sysname{} guides the target slicing with the prior knowledge from the buggy code, in an end-to-end way to pinpoint the slicing criterion in the targets. 
All feature slices are embedded and compared based on the vector similarity. 
Buggy candidates are audited to confirm unknown bugs in the targets. 
Experiments show that \sysname{} holds advantages in bug detection for real-world projects with acceptable efficiency. In total, \sysname{} has spotted 31 unknown bugs in the Linux kernel. All of them have been confirmed by the kernel developers, and 11 have been assigned CVEs. 
\end{abstract}


%\ccsdesc[500]{Security and privacy~Operating systems security}

%\begin{CCSXML}
%<ccs2012>
%  <concept_id>00000000.0000000.0000000</concept_id>
%  <concept_desc>Do Not Use This Code, Generate the Correct Terms for Your Paper</concept_desc>
%  <concept_significance>500</concept_significance>
%  <concept_id>00000000.00000000.00000000</concept_id>
%  <concept_desc>Do Not Use This Code, Generate the Correct Terms for Your Paper</concept_desc>
%  <concept_significance>300</concept_significance>
%  <concept_id>00000000.00000000.00000000</concept_id>
%  <concept_desc>Do Not Use This Code, Generate the Correct Terms for Your Paper</concept_desc>
%  <concept_significance>100</concept_significance>
%  <concept_id>00000000.00000000.00000000</concept_id>
%  <concept_desc>Do Not Use This Code, Generate the Correct Terms for Your Paper</concept_desc>
%  <concept_significance>100</concept_significance>
%</ccs2012>
%\end{CCSXML}

%\ccsdesc[500]{Do Not Use This Code~Generate the Correct Terms for Your Paper}
%\ccsdesc[300]{Do Not Use This Code~Generate the Correct Terms for Your Paper}
%\ccsdesc{Do Not Use This Code~Generate the Correct Terms for Your Paper}
%\ccsdesc[100]{Do Not Use This Code~Generate the Correct Terms for Your Paper}



\maketitle

\section{Introduction}
\label{sec:intro}

Detecting bugs through code similarity measurement has proven to be effective~\cite{kim2017vuddy,bowman2020vgraph,10.1145/3433210.3437533}. 
Generally, code elements (i.e., \textit{query}) related to a known bug are extracted and compared with the ones (i.e., \textit{targets}) obtained from a code base of interest. 
High similarity between the query and a target will indicate a potential bug in the corresponding target function. 

Directly measuring the similarity between functions is not always a good idea~\cite{zhang2022hunting}, as real-world functions usually contain not only the statements closely related to bugs but also many irrelevant ones that may differ significantly across functions. Such noise statements can heavily impede the function-level similarity detection. 

Recent studies tend to constrain the similarity measurement via various methods, but the core is to suppress the noise in the query via a slicing technique. 
For example, optimal vertex matching~\cite{zhang2022hunting} and subgraph isomorphism decision~\cite{gong2024sicode} adopt heavyweight matching methods, shrinking the query to contain only the bug-related elements instead of matching the entire target functions. 
Slice-based graph matching~\cite{DBLP:journals/tifs/HuangHYSLWW21} compares the similarity between slices related to input data, while FIRE~\cite{FIRE_usenix2024} matches buggy/patched slices with all target slices that are not purposefully selected according to the known bug. 

Though effective, the existing slice-based methods have explicit disadvantages. 
First, slicing only the query function is obviously affected by the noise in the target functions. In our experience, statement-by-statement matching can also emit lots of false matches~\cite{zhang2022hunting}, not to say a capability-restricted embedding model resolves a large target function to a small query slice~\cite{gong2024sicode}. 
Second, arbitrary slicing independent of the target bug can generate a large number of irrelevant slices. The influence of noise remains and probably affects the similarity measurement.

In this paper, we propose \sysname{} to overcome the noise in both the query and targets for precisely detecting bugs based on similarity measurement. 
The query is sliced based on identified bug elements by \sysname{} from the buggy/patched code. 
The targets are also sliced by \sysname{}, to ensure each target function is represented accurately by a slice that maximally resembles the query slice.
To achieve the goal, %arbitrary slicing must be discarded. 
\sysname{} leverages the knowledge of the known bug to direct the target slicing. 
More specifically, given a buggy function and its patched version, \sysname{} extracts the statements and variables that are tightly closed to the bug. 
Using them as prior knowledge, \sysname{} pinpoints the corresponding statements and variables in the target functions in an end-to-end way, with the help of similarity computation based on an encoder model. 
The pinpointed elements serve as the slicing criterion to guide target slicing. 
Because they are spotted via similarity measurement, the corresponding target slices are more likely to share similar semantic features with the query slice. 
Such slices are named \textit{feature slices} in this paper. 
Feature slices are embedded into vectors to ease the measurement of similarity between the query and targets. 
The top-ranked candidates will be manually audited to confirm unknown bugs.

We have implemented a prototype of \sysname{}. All embedding-related tasks utilize a fine-tuned pre-trained large code model, UniXcoder~\cite{guo-etal-2022-unixcoder}.
To improve efficiency, a coarse-grained function-level filtering is taken to screen out a small number of candidate functions from a large code base. 
\sysname{} has been proved to be effective and scalable in bug detection. 
In total, 31 previously unknown bugs have been detected in the Linux kernel v6.4-rc2 and subsequently confirmed by the kernel developers. 
Eleven of them are assigned CVEs. 
Over a benchmark with 76 bug pairs, our method demonstrates an advantage over nine matching-based methods in discovering bugs in large real-world projects. 

This paper makes the following contributions: 
\begin{itemize}
  
  \item 
  We propose a method to mitigate the noise in similarity-measurement-based bug detection. 
  The knowledge of a known bug is treated as a guide for pinpointing semantically similar code elements that are probably related to an unknown similar bug. 
  
  \item We propose a novel approach to obtaining the critical target elements end-to-end, based on the masked embeddings from an encoder model. It has been proven to be effective. 
  
  \item We design and implement \sysname{}, and evaluate it on the Linux kernel. Thirty-one previously unknown bugs have been detected. 
  Comparison with state-of-the-art (SOTA) tools demonstrates the effectiveness and efficiency of our method. 
  The artifacts are publicly available at \url{https://anonymous.4open.science/r/Yi3gA0k1aGz890}.
\end{itemize}

\section{Motivation}

\begin{figure*}[htbp]
  \centering
  \includegraphics[width=\linewidth]{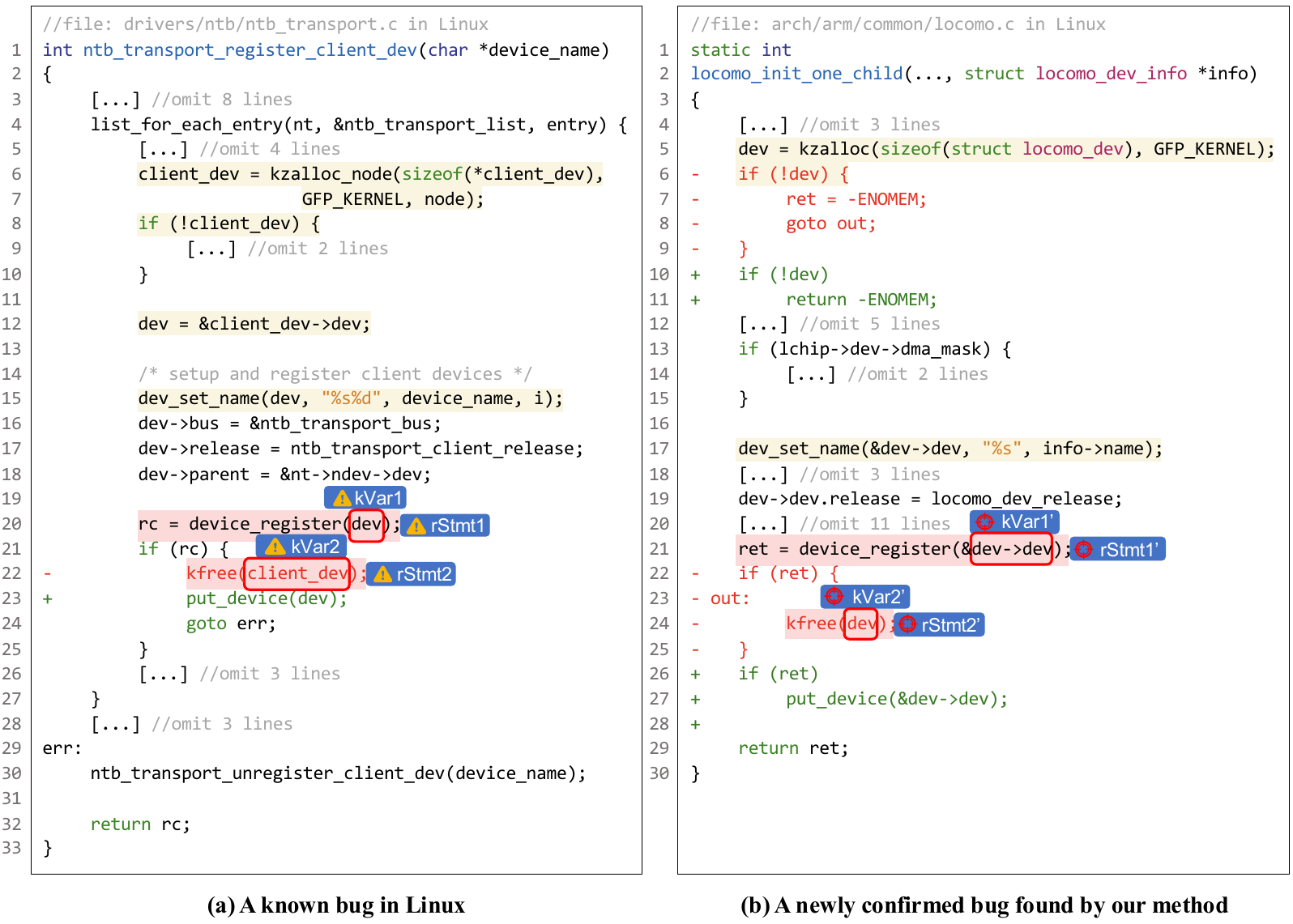}
  \caption{Motivating example. %(a) a known bug; 
  }
  \label{fig:motivating-example-1} % 这里的label用于之后引用图片
\end{figure*}

We use Figure~\ref{fig:motivating-example-1} to motivate our technique, which contains (a) a known bug and (b) a similar new bug in the Linux kernel.

Figure~\ref{fig:motivating-example-1}(a) is responsible for registering a list of NTB client devices with the transport layer. 
Each \textit{client\_dev} is allocated and initialized before being registered at Line 20. If any registration fails, the current device is destroyed to avoid memory leaks (Line 22), and all registered devices are cleaned up at Line 30. 
The bug arises when the device is simply destroyed with \texttt{kfree}, which only deallocates the memory by Line 6 but does not release the device name allocated by \texttt{dev\_set\_name} at Line 15, i.e., leading to a memory leak. 
The fix replaces \texttt{kfree} with \texttt{put\_device} (Line 23). The latter invokes a customized callback function \texttt{ntb\_transport\_client\_release} (registered at Line 17) to release all memory chunks associated with the device. 

In Figure~\ref{fig:motivating-example-1}(b), \texttt{locomo\_init\_one\_child} initializes and registers a single child device on the LoCoMo bus. 
% Similar with registering one client device in Figure~\ref{fig:motivating-example-1}(a), it exhibits the same memory leak bug and fix when the registration fails at Line 21. 
Similar to the process for registering a client device in Figure~\ref{fig:motivating-example-1}(a), this function contains the same memory leak bug if registration fails at Line 21, and the same fix applies. 
Identifying the bug alone would be challenging for auditors lacking sufficient domain expertise. 
Leveraging similarity-based matching techniques would help us recognize that the two functions implement similar semantics. 
However, the different code structures (loop \textit{v.s.} non-loop structures, continuous \textit{v.s.} intermittent operations on the devices) and device-specific operations (omitted lines) aggravate the function-level difference. 
The noise code makes the function difficult to stand out from hundreds of thousands of kernel functions when they are ranked based on their similarity to the left one. 
In fact, when we embed the functions into vectors with a fine-tuned code embedding model UniXcoder~\cite{guo-etal-2022-unixcoder} and compute the cosine similarity between vectors, \texttt{locomo\_init\_one\_child} is only ranked the 885-\textit{th} similar with \texttt{ntb\_transport\_register\_client\_dev}. 
Such a low ranking is far beyond the scope of a normal manual audit. 
To make the buggy function come into attention of the auditors, we need a method that can confidently enhance the similarity between the two functions and promote the ranking to a scope (e.g., within the top 10) acceptable to auditors. 

We present \sysname{} to highlight the bug based on similarity detection of bug-related slices that are embedded with pre-trained code models. 
According to the known bug and its fix, \sysname{} first identifies the variables (labelled as {kVar1} and {kVar2}) and the statements ({rStmt1} and {rStmt2}), which are immediately closely related to the bug. 
Then a customized slicing extracts other related statements, forming a feature slice (\textit{query}) of the bug, i.e., the highlighted lines in Figure~\ref{fig:motivating-example-1}(a). 
Facing the whole Linux kernel, \sysname{} leverages an encoder model to discover the counterpart variables and statements (i.e., the labelled ones in Figure~\ref{fig:motivating-example-1}(b)) in each target function. Target feature slices potentially possessing semantically similar operations with the query are extracted, e.g., the highlighted lines in Figure~\ref{fig:motivating-example-1}(b). 
An encoder model embeds all the feature slices, and \sysname{} ranks the target functions based on the cosine similarity between target slices and the query. 
By this means, Figure~\ref{fig:motivating-example-1}(b) is ranked first and can be quickly audited to identify the bug.

\section{Methodology}

\subsection{Overview}

\begin{figure*}[t]
  \centering
  \includegraphics[width=\linewidth,clip]{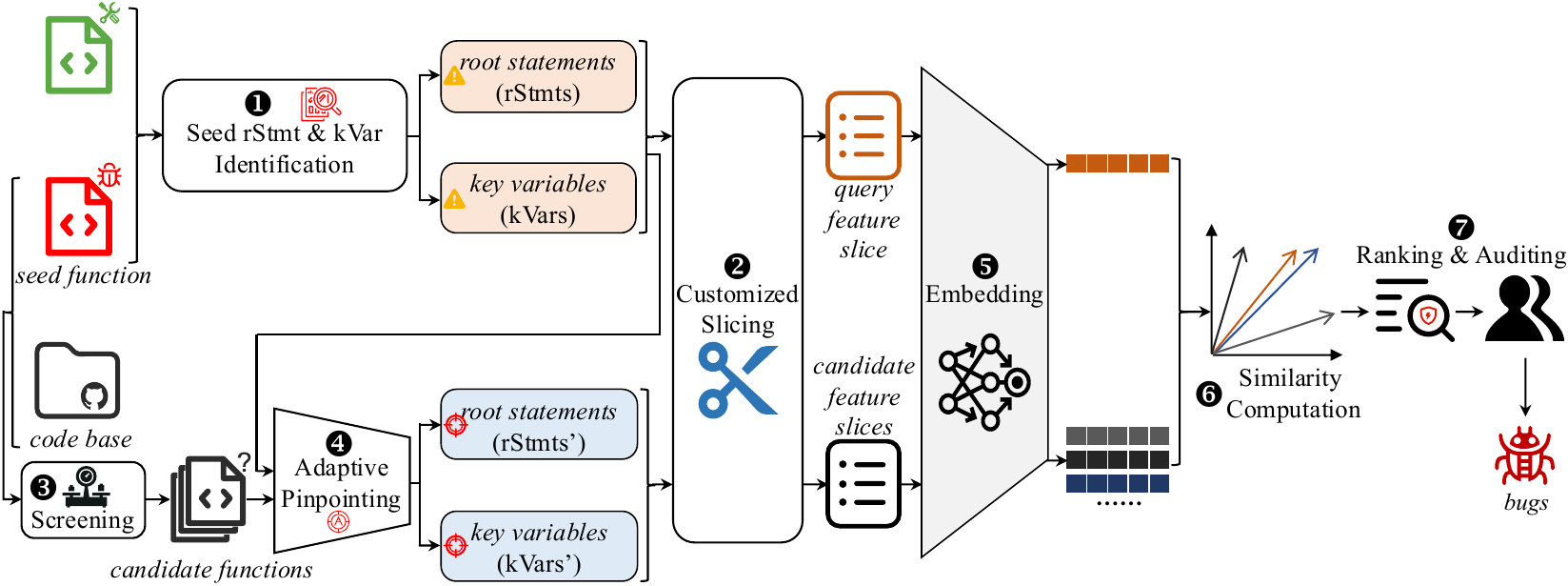}
  \caption{The overall workflow of \sysname{}.}
  \label{fig:approach_figure} 
\end{figure*}

Figure~\ref{fig:approach_figure} shows the workflow of \sysname{}. 
Given a seed function with a known bug and its fixed version, \sysname{} heuristically identifies two kinds of important elements (\ding{182}), i.e., \textit{key variables} and \textit{root statements}, which are then used to extract a customized slice as the \textit{query} (\ding{183}). We call the slice a \textit{feature slice} as it is highly related to and illustrates the feature of the bug. 
The target functions in the code base are filtered based on their coarse-grained similarity to the buggy seed function (\ding{184}), leaving a group of candidate functions for fine-grained similarity detection. 
With the help of an embedding model, the counterpart key variables and root statements in each candidate function are acquired (\ding{185}). Using them as the slicing criterion, the candidate feature slices are extracted (\ding{183}). 
All feature slices are embedded into vectors through an encoder model (\ding{186}), and the cosine similarity between each candidate slice and the query is computed (\ding{187}). 
All candidates are ranked according to the similarity for manual auditing to discover potential bugs (\ding{188}).

\subsection{Root Statements and Key Variables}
\label{section:RStmtKVar}

In this section, we informally define the root statements and key variables that are critical in Figure~\ref{fig:approach_figure} for extracting the query feature slice. The ones in the seed function also guide the discovery of the slicing criterion in candidate functions. 

We define a \textit{root statement} (shortened as \textit{rStmt}) as a statement that can immediately trigger the bug in the buggy code. 
It involves two cases. First, the statement itself is buggy and hence modified or removed in the fixed version. 
Line 22 in Figure~\ref{fig:motivating-example-1}(a) is an example, which partially releases the allocated memory, resulting in a leak. 
Second, the statement immediately reaches the bug point through the control flow when a certain condition is satisfied. 
Line 20 in Figure~\ref{fig:motivating-example-1}(a) is an example. If the device registration fails, the immediate error handling runs into a bug. 

A \textit{key variable} (shortened as \textit{kVar}) is an operand in the root statement, which typically holds an object or a value that is highly related to the causation of the bug. 
In this study, it can be either a simple variable or a compound expression, as an argument or the resultant value. 
In Figure~\ref{fig:motivating-example-1}(a), \textit{client\_dev} and \textit{dev} are two key variables, but \textit{rc} is not considered such because it only indicates a condition and does not involve an object that the buggy flow operates. 

In a target (or candidate) function, we refer to the elements pinpointed using the seed rStmts and kVars as candidate rStmts and kVars, as they may exhibit semantic similarity to their seed counterparts and be involved in potentially similar bugs.

\subsection{Identifying Seed Root Statements and Key Variables}
\label{section:seedRstmtKvar}

According to Figure~\ref{fig:approach_figure}, identifying the seed rStmts and kVars is a critical step. 
Section~\ref{section:RStmtKVar} has presented a conceptual guide to the identification, but it is not suitable for implementing an automated tool. For example, a skilled human reviewer may understand why some variables are unnecessary to be kept as kVars, but the sense cannot be well automated. %Therefore, in this section, 
To enhance the efficiency of automated identification, we adopt an automated approach inspired by~\cite{VulDetector2020TIFS} %, which recognizes the kVars first and then the rStmts. 
to recognize the kVars and then heuristically pinpoint the rStmts. 

\textbf{Key variable identification.}
We adopt the term frequency~\cite{VulDetector2020TIFS} to calculate the importance of potential kVars, which measures the frequency of a variable's occurrence within the patch. A higher count indicates the variable is more critical to the bug. 
Any addition, deletion or modification will increase the frequency of the involved variables. 
The ones with the highest frequency and existing in the buggy code will be treated as kVars. 
When more than one variable reaches the same highest frequency, all of them are taken into account. 
Note that, if a variable appears in the \textit{return} statement, either in the buggy or the fixed code, it is eliminated from the candidates. Such a rule is included because we find that in large code bases like the Linux kernel, a bug patch may contain multiple similar blocks for error handling and every block returns the same variable indicating a failure of the execution. 
Figure~\ref{fig:motivating-example-1}(a) is a perfect example. We can easily highlight \textit{dev} and \textit{client\_dev} as two kVars as there are only two variables in the patch (Lines 22 $\thicksim$ 23) and both have the same frequency (i.e., 1). 

\textbf{Root statement screening.}
Based on the identified kVars, we define two rules, corresponding to the two cases in Section~\ref{section:RStmtKVar}, to collect the rStmts starting from the patch statements. %from the buggy code. 

First, if a kVar $\textit{var}_\textit{k}$ appears in a deleted/modified statement, it is directly collected, e.g., Line 22 in Figure~\ref{fig:motivating-example-1}(a). 
Second, if $\textit{var}_\textit{k}$ appears in an inserted statement that is not in the buggy code, 
we find another correlated statement $\textit{stmt}_\textit{r}$ as the corresponding rStmt. 
Our heuristic picks $\textit{stmt}_\textit{r}$ that embodies contains $\textit{var}_\textit{k}$ and appears before and in the closest proximity to the inserted statement. 
Take Figure~\ref{fig:motivating-example-1}(a) as an example. With $\textit{var}_\textit{k} =$ \textit{dev}, we can determine that $\textit{stmt}_\textit{r}$ = Line 20 which uses \textit{dev} and has only one step to Line 21.

\subsection{Pinpointing Candidate Root Statements and Key Variables}
\label{section:pinpointing}

Candidate rStmts and kVars in target functions are crucial to extract feature slices that are potentially semantically similar to the query. 
Provided a seed rStmt $\textit{s}_\textit{stmt}$ and kVar $\textit{s}_\textit{var}$, 
we propose an end-to-end approach to identifying the candidate rStmt $\textit{c}_\textit{stmt}$ and kVar $\textit{c}_\textit{var}$ simultaneously in a target function \textit{F}. 
The approach harnesses the power of a transformer-based encoder model, which learns the embedding representation of each token by taking all other tokens in the input into consideration. 
In other words, the embedding of a specific token has already comprised the context information. 
Inspired by it, our approach determines $\textit{c}_\textit{stmt}$ and $\textit{c}_\textit{var}$ in $F$ with Equation~\ref{eq:pinpointCandidateRstmtKvar}, 
where the $\arg\max$ is taken over all pairs $(\textit{stmt}_{\textit{i}}, \textit{var}_{\textit{j}})$. Each pair contains a statement $\textit{stmt}_{\textit{i}}$ in $F$ and a variable occurrence $\textit{var}_{\textit{j}}$ in the statement. 
%Here, 
$\mathbf{v}_\textit{x}$ is the embedding of $\textit{x}$, and $\textit{sim}()$ computes vector similarity. 

\begin{equation}
    \textit{c}_{\textit{stmt}},~ \textit{c}_{\textit{var}} = \underset{\forall (\textit{stmt}_{\textit{i}}, \textit{var}_{\textit{j}}) }{\arg\max} \textit{sim}(\mathbf{v}_{\textit{s}_\textit{var}}, \mathbf{v}_{\textit{var}_{\textit{j}}})
    \label{eq:pinpointCandidateRstmtKvar}
\end{equation}

Note that, each $\textit{var}_\textit{j}$ is constrained by $\textit{stmt}_\textit{i}$ where the variable appears. That means, if a variable appears in different statements, it will possess different embeddings associated with the occurrences. 
Moreover, as the position matters in transformer-based encoder models, a variable occurring twice in a statement will emit two different embeddings. 
We can leverage the characteristics to distinguish $\textit{c}_\textit{var}$ from all the other occurrences of the same variable, and at the same time pinpoint the corresponding $\textit{c}_\textit{stmt}$. 

There are two ways to obtain the embedding of a specific \textit{var} in a \textit{stmt}, as shown in Equations~\ref{eq:concatTokens} and~\ref{eq:maskBased}, where %`T' is the tokenizer used by the encoder model. 
Tok: $C \rightarrow T$ is the tokenizer that splits a given text $C$ into a sequence of tokens $T$ and Enc: $T \rightarrow (R^d)^n$ is the encoder model which encodes each among the $n$ tokens into a $d$-dimensional floating vector.

\begin{equation}
    \mathbf{v}_\textit{var} = \sum_{\textit{j} = \textit{i}}^{\textit{k}} \mathbf{v}_{\textit{t}_\textit{j}},\,{\rm with }\, {\rm Enc(Tok}(\textit{stmt})) \rightarrow \cdots, \mathbf{v}_{\textit{t}_\textit{j}}, \cdots\,\,{\bf and }\,\, {\rm Tok}(\textit{var}) \rightarrow \textit{t}_\textit{i}, \cdots, \textit{t}_\textit{k}
    \label{eq:concatTokens} 
\end{equation}

\begin{equation}
    \mathbf{v}_\textit{var} = {\mathbf{v}_{\texttt{[MASK]}}}, \,{\rm with}\, {\rm Enc}(\cdots, \textit{t}_{\textit{i}-1}, \texttt{[MASK]}, \textit{t}_{\textit{k}+1}, \cdots) \rightarrow \cdots,  \mathbf{v}_{\textit{t}_{\textit{i}-1}}, \mathbf{v}_{\texttt{[MASK]}}, \mathbf{v}_{\textit{t}_{\textit{k}+1}}, \cdots
    % \mathbf{v}_{var} = \underset{\cdots, \underbrace{\scalebox{.75}{\texttt{[MASK]}}}_{var}, \cdots = {\rm tokenize}(stmt)}{\mathbf{v}_{\texttt{[MASK]}}}
    \label{eq:maskBased}
\end{equation}

Equation~\ref{eq:concatTokens} aggregates the embeddings of the tokens corresponding to \textit{var}, utilizing both the context information and the lexical components of the variable. 
Equation~\ref{eq:maskBased} replaces the variable with a special token \texttt{[MASK]}, %\semihide{ (i.e., $\frac{\texttt{[MASK]}}{\cancel{\textit{var}}}$)}, 
leveraging the filling-the-blank ability of the model to infer the representation of \textit{var}. 
Surprisingly, the lexical information does not provide much help and in Section~\ref{section:ablation:choices}, the advantage of the second method is illustrated compared to the first.
In this study, we take Equation~\ref{eq:maskBased} to compute the variable embeddings for both the seed and the candidate kVars.

Considering that we aim to obtain a slice with multiple statements that can represent potential error-prone semantics of some objects of interest, \sysname{} excludes the variables occurring only once (apart from the declarations) throughout the given function, i.e., they are not considered for candidate kVars and hence not masked in this step.

% \begin{figure*}[htbp]
\begin{figure}[t]
  \centering
  \includegraphics[width=\linewidth,clip,]{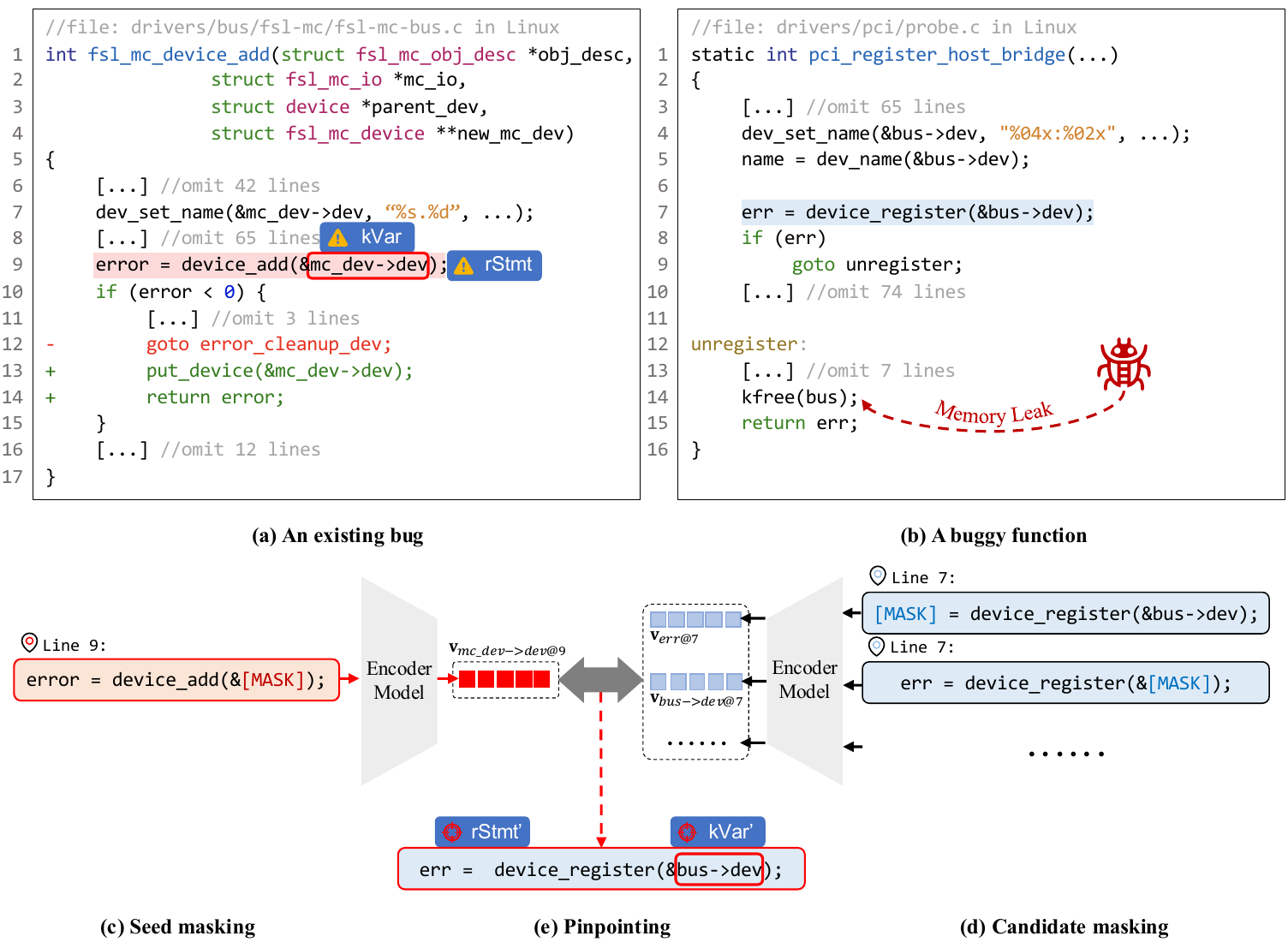}
  \caption{
  An example illustrating how to pinpoint the candidate root statement and key variable in a candidate function based on their seed counterparts.
  }
  \label{fig:mask_example_figure} % 这里的label用于之后引用图片
  \vspace{-5pt} 
\end{figure}

\textbf{Example.} 
Figure~\ref{fig:mask_example_figure} illustrates the workflow. 
With the identified rStmt and kVar in (a), \sysname{} masks the kVar and obtains its embedding $\mathbf{v}_\textit{kv}$ (in (c)). 
For the variables in a candidate function as in (b), we mask their occurrences one by one. 
Take Line 7 in (b) as an example, which contains two variables. We build two masked statements as (d) shows, each with only one \texttt{[MASK]} corresponding to a specific variable, and compute the variables' embeddings, i.e., $\mathbf{v}_{\textit{err}@7}$ and $\mathbf{v}_{\textit{bus\scalebox{.9}{\tt ->}dev}@7}$. 
Applying Equation~\ref{eq:pinpointCandidateRstmtKvar} to all the variables' embeddings, \sysname{} determines the candidate rStmt and kVar as in (e), i.e., Line 7 and \textit{bus\scalebox{.9}{\tt ->}dev} in (b), respectively. 
Following~\cite{lukangjie_CCS22, lukangjie_SEC24}, we consider \textit{bus\scalebox{.9}{\tt ->}dev} as a variable.

\subsection{Customized Slicing}
\label{section:slicing}

%\SetInd{1ex}{0.4em}
\SetInd{1ex}{2em}
\begin{algorithm2e}
\caption{Customized slicing}
\label{alg:slicing}
\newcommand{\var}[1]{\textit{#1}}
\newcommand{\call}[1]{\textit{#1}}
\SetKwFunction{slice}{\rm \textit{customizedSlice}}
\SetKwFunction{collect}{\rm \textit{collectAndPrepare}}
\DontPrintSemicolon
\newcommand\mycommfont[1]{\small\textnormal{#1}}
\SetCommentSty{mycommfont}
\SetKwComment{tcp}{$\triangleright$ }{}
\SetKwComment{tcc}{$\triangleright$ }{}
\SetKwProg{Fn}{Func}{:}{}%

\Fn{\slice{F, rStmt, kVar}}{
    \var{kVar.depth} $\gets$ 0 \\ %\textbf{;} 
    $\varSigma \gets [\var{rStmt}]$ \tcp*[r]{collected stmts} %\textbf{;} 
    $\varGamma \gets [\var{kVar}]$  \tcp*[r]{tracked variables}
    $\var{cfg}, \var{ddg}  \gets \call{buildCfgAndDdg}(\var{F})$ \label{alg:line:buildcfgddg} \\
    % $\var{P} \gets \var{cfg}.\call{getLongestPathStartsFrom}(\var{rStmt})$ \label{alg:line:getlongest} \\
    \While{$(\gamma \gets \varGamma.\call{removeFirst}())$ {\rm is not} \textit{nil}}{ \label{alg:line:while}
        \lIf(\tcp*[f]{limit slicing depth}){$\gamma.\var{depth} > 1$}{ \label{alg:line:check depth}
            \textbf{continue}
        }
        $\var{v}, \var{stmt} \gets \var{ddg}.\call{getDefinition}(\gamma)$ \label{alg:line:getdef} \\
        % $\underline{\call{collectAndPrepare}}(\gamma, \var{stmt}, \varSigma, \varGamma)$ \tcp*[r]{Lines~\ref{alg:line:collect}/\ref{alg:line:cond define}} \label{alg:line:onestep 1} 
        $\underline{\call{collectAndPrepare}}(\gamma, \var{stmt}, \varSigma, \varGamma)$ \label{alg:line:onestep 1} \tcp*[r]{collect stmt @\ref{alg:line:collect} and var @\ref{alg:line:gamma append 1}}
        
        $\varPsi \gets \var{ddg}.\call{getAllUses}(\var{v})$ \tcp*[r]{all use stmts}  \label{alg:line:getuses} %% comment right aligned, no need a '\\'
        \For{$\psi \in \varPsi$}{ \label{alg:line:slice}
            \uIf{$\var{cfg}.\call{hasFwdPath}(\psi, \var{rStmt})$}{ \label{alg:line:earlier}
                % $\underline{\call{collectAndPrepare}}(\var{v}, \psi, \varSigma, \varGamma)$ \tcp*[r]{Lines~\ref{alg:line:collect}/\ref{alg:line:cond use}} \label{alg:line:onestep 2}
                $\underline{\call{collectAndPrepare}}(\var{v}, \psi, \varSigma, \varGamma)$ \label{alg:line:onestep 2} \tcp*[r]{collect stmt @\ref{alg:line:collect} and var @\ref{alg:line:gamma append 2}}
            }
            \uElseIf{$\var{cfg}.\call{hasFwdPath}(\var{rStmt}, \psi)$}{ \label{alg:line:after}
                % $\underline{\call{collectAndPrepare}}(\var{v}, \psi, \varSigma, \varGamma)$ \tcp*[r]{Lines~\ref{alg:line:collect}/\ref{alg:line:cond use}} \label{alg:line:onestep 3}
                $\underline{\call{collectAndPrepare}}(\var{v}, \psi, \varSigma, \varGamma)$ \label{alg:line:onestep 3} \tcp*[r]{collect stmt @\ref{alg:line:collect} and var @\ref{alg:line:gamma append 2}}
            }
        }
    } \label{alg:line:while end}
    $\varSigma \gets \call{filterByMaxCoveragePath}(\var{cfg}, \var{rStmt}, \varSigma)$  \label{alg:line:max filter} \\
    \KwRet $\call{sortByLineNumber}(\varSigma, \var{deduplicate} = \textsc{True})$ \label{alg:line:return}
}
\Fn{\collect{var, stmt, $\varSigma$, $\varGamma$}}{
    \lIf{ $\call{isNormalStmt}(\var{stmt}) == \textsc{False}$}{ \KwRet }
    $\varSigma.\call{append}(\var{stmt})$ \tcp*[r]{collect \var{stmt}} \label{alg:line:collect} 
    $\var{l} \gets \call{getLhs}(\var{stmt})$ \tcp*[r]{\var{l} == \var{v}@{\ref{alg:line:getdef}}} \label{alg:line:getlhs} 
    $\var{R} \gets \call{getRhsOperands}(\var{stmt})$  \label{alg:line:getrhs} \\
    \uIf(\tcp*[f]{unary op}){\var{l} \rm{is not} \var{nil} \textbf{and} $\call{len}(\var{R}) == 1$}{ \label{alg:line:unary}
            \uIf(\tcp*[f]{define \var{var}, i.e., \var{v}@\ref{alg:line:getdef} == {$\gamma$}@\ref{alg:line:getdef}}){\var{l} == \var{var}}{ \label{alg:line:cond define}
                $\var{R}[0].\var{depth} \gets \var{var.depth} + 1$ \label{alg:line:update depth 1} \\
                $\varGamma.\call{append}(\var{R}[0])$ \label{alg:line:gamma append 1}  \tcp*[r]{prepare one more step for \var{R}[0]}
            }
            \uElse(\tcp*[f]{use \var{var} and \var{var} == \var{R}[0]}){  \label{alg:line:cond use}
                $\var{l.depth} \gets \var{var.depth} + 1$ \label{alg:line:update depth 2} \\
                $\varGamma.\call{append}(\var{l})$ \label{alg:line:gamma append 2}  \tcp*[r]{prepare one more step for \var{l}}
            }
    } \label{alg:line:unary end}
}
\end{algorithm2e}

Instead of applying a traditional slicing algorithm to collect all the reachable statements in the dependency graph of function \textit{F} with the slicing criterion of the identified rStmt and kVar, we customize the slicing to constrain the resultant slice to be tightly related to the (potential) bug. 
The customization is reflected in two aspects, as shown in Algorithm~\ref{alg:slicing}. 

First, we limit the reachable depth in the dependency graph. Only when the kVar is the definition (solo use) of a unary expression, will one more step be forwarded to track the corresponding use (definition) variable (Lines~\ref{alg:line:unary} $\thicksim$~\ref{alg:line:unary end}). 
That is why \textit{rc} and its correlated lines are not included in the slice for the kVar \textit{dev} in Figure~\ref{fig:motivating-example-1}. 
Otherwise, only the statements directly depending on the kVar will be collected. The constraint is enforced by binding a \textit{depth} to the variables, which are checked for each tracked variable (Line~\ref{alg:line:check depth}) and updated for a potentially trackable variable (Lines~\ref{alg:line:update depth 1} and~\ref{alg:line:update depth 2}).

Second, statements referring to the trackable variables are not all kept in the slice. 
Instead, we only retain the statements in the control flow path that can cover the rStmt and most of the trackable variables. The strategy makes the slice compact and simultaneously keeps the semantics of the involved objects as far as possible in the slice. 
Line~\ref{alg:line:max filter} filters the result collected through the while loop (Lines~\ref{alg:line:while} $\thicksim$~\ref{alg:line:while end}).

It is also notable that we treat a member access expression in the form of \textit{mc\_dev\texttt{->}dev} in Figure~\ref{fig:mask_example_figure}(a) as a whole, as is done in~\cite{lukangjie_CCS22, lukangjie_SEC24}, and ignore the access to any member from a tracked variable via the `\texttt{->}' operator. 
Neither do we track the receiver pointer (e.g., \textit{mc\_dev}) individually in such cases. 
The rule applies to both variable gathering and dependency analysis  (Lines~\ref{alg:line:getdef},~\ref{alg:line:getuses},~\ref{alg:line:getlhs} and~\ref{alg:line:getrhs}). 
In other words, \textit{v} is literally equal to $\gamma$ at Line~\ref{alg:line:getdef}.

All the statements of interest are maintained in $\varSigma$, which is sorted by the line numbers before the algorithm returns it as the resulting slice (Line~\ref{alg:line:return}). 
The duplicate statements are also refined along with the sorting. 
Furthermore, if more than one rStmt and kVar are identified for a function, their slices are combined to compose the final feature slice.

\subsection{Embedding, Ranking and Auditing}
\label{section:embedding ranking}

All feature slices are embedded into vectors with a pre-trained encoder model. 
In this study, we choose UniXcoder~\cite{guo-etal-2022-unixcoder}, a unified cross-modal model for code representation, as the embedding model. 
Since it learns from multiple languages but we focus on C code, we fine-tune the model following its instruction on the C-based code clone dataset POJ-104~\cite{lu2021codexglue}. 
The fine-tuned model is termed UniXcoder-POJ in this paper. 

Cosine similarity is computed between the query and each candidate feature slice. The candidate functions are then ranked based on the similarity values in descending order. 
The results are manually audited to confirm unknown bugs. 
Following previous studies~\cite{zhang2022hunting, gong2024sicode} that audit the top-ranked functions, in this study, we select the top 10 candidates for auditing.

\subsection{Function-level Candidate Function Screening}
\label{section:Screening}

In large code bases like the Linux kernel, directly pinpointing candidate rStmts and kVars and then computing the similarity between their corresponding slices and the query will inevitably encounter the efficiency issue. 
To address the issue, we propose a function-level screening step to screen out a set of candidate functions from the whole code base (\ding{184} in Figure~\ref{fig:approach_figure}). 
Functions are embedded into vectors, and similarity between each target and the seed function is computed. The top similar functions are left for further analysis, i.e., pinpointing candidate rStmts and kVars. 
In this study, we choose to keep the top 1000 functions as the candidates.

\section{Evaluation}

\subsection{Experiment Setup}
\label{section:exp setup}

We have implemented a prototype of \sysname{}. 
Code preprocessing and slicing are built on top of Joern~\cite{joern_io_Joern_The_Bug_2024} and fuzzyc2cpg~\cite{fuzzyc2cpg}, while the other components (including model tuning, embedding and matching) are implemented in Python. 
All experiments are conducted on a server with Intel(R) Xeon(R) Gold 5218 CPU@2.30GHz, 256 GB memory, Ubuntu 20.04 and NVIDIA GeForce RTX 3090 GPU (for fine-tuning, embedding and similarity calculation).

\textbf{Datasets.}
\label{section:Dataset}
We evaluate \sysname{} on two datasets. First, we select the Linux kernel (shortened as \textit{Linux}) v6.4-rc2 as the target of evaluation for detecting new bugs and assessing the efficiency. 
To improve the matching efficiency, we precompute the function embeddings, slices, their corresponding embeddings, and the \texttt{[MASK]} embeddings (see Figure~\ref{fig:mask_example_figure}(d)) offline in a one-time preprocessing phase, leveraging their reusable nature. 
In total, 293,292 functions are successfully extracted 
and {10,909,165} slices are obtained. 
Second, we construct a benchmark with 76 pairs of seed/target bugs. The benchmark consists of (D1) the 31 bug pairs by \sysname{} from Linux v6.4-rc2, (D2) 31 collected from the Linux commit history at the inception of this study, and (D3) 14 from OpenSSL v3.0.8 and Linux v6.2.5 by SICode~\cite{gong2024sicode}. The dataset is publicly available at the URL provided in Section~\ref{sec:intro}. 

\textbf{Comparative methods.}
We compare \sysname{} with nine publicly available, open-source methods 
that measure the code similarity, including ReDeBug~\cite{jang2012redebug}, SourcererCC~\cite{sajnani2016sourcerercc}, Infercode~\cite{Infercode2021}, NIL~\cite{NIL2021}, VUDDY~\cite{kim2017vuddy}, FIRE~\cite{FIRE_usenix2024}, SICode~\cite{gong2024sicode}, UniXcoder~\cite{guo-etal-2022-unixcoder} and UniXcoder-POJ. 
The first four methods target discovering cloned functions; VUDDY, FIRE and SICode detect bugs in similar implementations; and the last two can be used as clone detectors and also serve as the embedding model in \sysname{}. 
Unless specified otherwise, we manually audit the top 10 reported instances for Infercode, SICode and UniXcoder models where the ranking is available, or the reported candidates for VUDDY, ReDeBug, SourcererCC, NIL and FIRE that emit found/not-found results. 
In addition, we inspect the rankings of confirmed bugs across all tools. 
Specifically, the comparative tools are summarized as follows.

\begin{itemize} 

    \item ReDeBug~\cite{jang2012redebug}: a fast, syntax-driven tool for detecting unpatched code clones in large, multi-language code bases, excelling in scale and speed. 

    \item SourcererCC~\cite{sajnani2016sourcerercc}: a token-based code clone detector using an optimized inverted-index and filtering heuristics. 
    
    \item Infercode~\cite{Infercode2021}: a self-supervised learning approach for generating AST-based code embeddings, effective in retrieving similar functions. 

    \item NIL~\cite{NIL2021}: SOTA token-based clone detector, using N-grams and inverted indexes to identify large variations in large code bases. 
    
    \item VUDDY~\cite{kim2017vuddy}: a scalable and efficient tool for detecting vulnerable code clones, focusing on function-level granularity. 
    
    \item FIRE~\cite{FIRE_usenix2024}: SOTA similarity-based bug detector, combining {multi-stage filtering} and {differential taint path analysis} to achieve efficient and precise detection. 
    
    \item SICode~\cite{gong2024sicode}: SOTA embedding-based bug detector, identifying subgraph isomorphism based on code graph embeddings to detect bugs. 
    
    \item UniXcoder~\cite{guo-etal-2022-unixcoder} and UniXcoder-POJ: cross-modal pre-trained model designed for programming languages that enhances code representation with AST and comments, excelling in a range of code-related tasks. UniXcoder-POJ is a fine-tuned model (see Section~\ref{section:embedding ranking}).

\end{itemize}

The tools are executed with their default parameter setting, 
but we increase the maximum token sequence length from 512 to 1024 in the use of UniXcoder models. 
Functions exceeding the length are truncated, but statistics show 92.61\% of the functions are kept unchanged. 
The dimension of embedding vectors is 768 in \sysname{}.

% \usepackage{multirow}
% \begin{table}[htbp]
\begin{table}[t]
\centering
\caption{%Bug detection results
Newly confirmed bugs in Linux v6.4-rc2.
}
\label{tab:table_all_results}
\setlength{\tabcolsep}{3pt}
%\small
%\scriptsize
%\tiny
\begin{tabular}{|c|c|c|c|c|c|c|c|c|c|c|c|}
%\begin{tabularx}{\linewidth}{|X|X|X|X|X|X|X|X|X|X|X|X|}
%\begin{tabularx}{\linewidth}{|c|c|c|c|c|c|c|c|c|c|c|c|}
\hline
\textbf{ID} & 
  \textbf{Seed / Detected Buggy Functions} &
  \rotatebox{90}{\textbf{ReDeBug}} &
  \rotatebox{90}{\textbf{SourcererCC}} &
  \rotatebox{90}{\textbf{Infercode}} &
  \rotatebox{90}{\textbf{NIL}} &
  \rotatebox{90}{\textbf{VUDDY}} &
  \rotatebox{90}{\textbf{FIRE}} &
  \rotatebox{90}{\textbf{SICode}} &
  \rotatebox{90}{\textbf{UniXcoder}} &
  \rotatebox{90}{\textbf{UniXcoder-POJ}} &
  \rotatebox{90}{\textbf{\sysname{}}} \\ \hline
1           & \textit{\scriptsize mlx5e\_ipsec\_remove\_trailer}                   / \textit{\scriptsize esp\_remove\_trailer(ipv6)}             & \(\times\)               & \(\times\)                   & \(\times\)                 & \(\times\)           & \(\times\)             & \(\times\)            & \(\times\)            & 4                  & 4                      & 3              \\ \hline
2           & \textit{\scriptsize vb2\_dc\_put\_userptr}                           / \textit{\scriptsize vb2\_vmalloc\_put\_userptr}             & \(\times\)               & \(\times\)                   & \(\times\)                 & \(\times\)           & \(\times\)             & \(\times\)            & \(\times\)            & 11                 & 2                      & 1              \\ \hline
3           & \textit{\scriptsize radeon\_tv\_get\_modes}                          / \textit{\scriptsize amdgpu\_vkms\_conn\_get\_modes}         & \(\times\)               & \(\times\)                   & \(\times\)                 & \(\times\)           & \(\times\)             & \(\times\)            & \(\times\)            & 18                 & {\small 168}                    & 2              \\ \hline
4           & \textit{\scriptsize ath11k\_update\_per\_peer\_tx\_stats}            / \textit{\scriptsize ath12k\_update\_per\_peer\_tx\_stats}   & \(\times\)               & \(\checkmark\)                   & 2                  & \(\checkmark\)           & \(\times\)             & \(\checkmark\)            & \(\times\)            & 2                  & 2                      & 1              \\ \hline
5           & \textit{\scriptsize pdsc\_auxbus\_dev\_register}                     / \textit{\scriptsize add\_adev}                              & \(\times\)               & \(\times\)                   & \(\times\)                 & \(\times\)           & \(\times\)             & \(\times\)            & 6            & 33                 & {\small 530}                    & 1              \\ \hline
6           & \textit{\scriptsize amdgpu\_vkms\_conn\_get\_modes}                  / \textit{\scriptsize nv17\_tv\_get\_hd\_modes}               & \(\times\)               & \(\times\)                   & \(\times\)                 & \(\times\)           & \(\times\)             & \(\times\)            & \(\times\)            & 19                 & 23                     & 3              \\ \hline
7           & \textit{\scriptsize amdgpu\_vkms\_conn\_get\_modes}                  / \textit{\scriptsize radeon\_add\_common\_modes}             & \(\times\)               & \(\times\)                   & \(\times\)                 & \(\checkmark\)           & \(\times\)             & \(\times\)            & \(\times\)            & 3                  & 3                      & 1              \\ \hline
8           & \textit{\scriptsize nouveau\_connector\_get\_modes}                  / \textit{\scriptsize nv17\_tv\_get\_ld\_modes}               & \(\times\)               & \(\times\)                   & \(\times\)                 & \(\times\)           & \(\times\)             & \(\times\)            & \(\times\)            & 83                 & {\small 654}                    & 1              \\ \hline
9           & \textit{\scriptsize probe\_uprobe\_multi\_link}                      / \textit{\scriptsize serial\_test\_fexit\_stress}            & \(\times\)               & \(\times\)                   & \(\times\)                 & \(\times\)           & \(\times\)             & \(\times\)            & \(\times\)            & 38                 & 10                     & 2              \\ \hline
10          & \textit{\scriptsize ath11k\_mhi\_register}                           / \textit{\scriptsize ath12k\_mhi\_register}                  & \(\times\)               & \(\times\)                   & 29                 & \(\times\)           & \(\times\)             & \(\times\)            & \(\times\)            & 3                  & 2                      & 1              \\ \hline
11          & \textit{\scriptsize tpg110\_get\_modes}                              / \textit{\scriptsize psb\_intel\_lvds\_get\_modes}           & \(\times\)               & \(\times\)                   & \(\times\)                 & \(\times\)           & \(\times\)             & \(\times\)            & \(\times\)            & 75                 & {\small 126}                    & 1              \\ \hline
12          & \textit{\scriptsize aldebaran\_mode2\_suspend\_ip}                   / \textit{\scriptsize smu\_v13\_0\_10\_mode2\_suspend\_ip}    & \(\times\)               & \(\checkmark\)                   & 3                  & \(\checkmark\)           & \(\times\)             & \(\checkmark\)            & \(\times\)             & 3                  & 3                      & 1              \\ \hline
13          & \textit{\scriptsize dm9601\_mdio\_read}                              / \textit{\scriptsize sr\_mdio\_read}                         & \(\times\)               & \(\times\)                   & \(\times\)                 & \(\checkmark\)           & \(\times\)             & \(\times\)            & \(\times\)            & 6                  & 2                      & 3              \\ \hline
14          & \textit{\scriptsize imx8mp\_blk\_ctrl\_probe}                        / \textit{\scriptsize imx8m\_blk\_ctrl\_probe}                & \(\times\)               & \(\checkmark\)                   & \(\times\)                 & \(\checkmark\)           & \(\times\)             & \(\checkmark\)            & \(\times\)            & 3                  & 2                      & 1              \\ \hline
15          & \textit{\scriptsize amdgpu\_vkms\_prepare\_fb}                       / \textit{\scriptsize amdgpu\_dm\_plane\_helper\_prepare\_fb} & \(\times\)               & \(\times\)                   & \(\times\)                 & \(\times\)           & \(\times\)             & \(\times\)            & \(\times\)            & 2                  & 2                      & 1              \\ \hline
16          & \textit{\scriptsize emif\_get\_id}                                   / \textit{\scriptsize \_emif\_get\_id}                        & \(\times\)               & \(\times\)                   & 2                  & \(\checkmark\)           & \(\times\)             & \(\checkmark\)            & \(\times\)            & 2                  & 2                      & 1              \\ \hline
17          & \textit{\scriptsize register\_device}                                / \textit{\scriptsize parport\_attach}                        & \(\times\)               & \(\times\)                   & \(\times\)                 & \(\times\)           & \(\times\)             & \(\times\)            & \(\times\)            & {\small 113}                & 23                     & 2              \\ \hline
18          & \textit{\scriptsize mt7915\_thermal\_init}                           / \textit{\scriptsize mt7921\_thermal\_init}                  & \(\times\)               & \(\times\)                   & \(\times\)                 & \(\checkmark\)           & \(\times\)             & \(\times\)            & \(\times\)            & 2                  & 2                      & 1              \\ \hline
19          & \textit{\scriptsize mt7915\_thermal\_init}                           / \textit{\scriptsize mt7615\_thermal\_init}                  & \(\times\)               & \(\times\)                   & \(\times\)                 & \(\times\)           & \(\times\)             & \(\times\)            & \(\times\)            & 3                  & 3                      & 1              \\ \hline
20          & \textit{\scriptsize dcn32\_enable\_phantom\_plane}                   / \textit{\scriptsize enable\_phantom\_plane}                 & \(\times\)               & \(\checkmark\)                   & 1                  & \(\checkmark\)           & \(\times\)             & \(\checkmark\)            & \(\times\)            & 2                  & 2                      & 1              \\ \hline
21          & \textit{\scriptsize rt5682s\_register\_dai\_clks}                    / \textit{\scriptsize rt5682\_register\_dai\_clks}            & \(\times\)               & \(\times\)                   & \(\times\)                 & \(\checkmark\)           & \(\times\)             & \(\checkmark\)            & 3            & 2                  & 2                      & 1              \\ \hline
22          & \textit{\scriptsize esp\_remove\_trailer(ipv6)}                      / \textit{\scriptsize esp\_remove\_trailer(ipv4)}             & \(\times\)               & \(\checkmark\)                   & 2                  & \(\checkmark\)           & \(\times\)             & \(\checkmark\)            & 1            & 2                  & 2                      & 1              \\ \hline
23          & \textit{\scriptsize i2c\_register\_adapter}                          / \textit{\scriptsize usb\_new\_device}                       & \(\times\)               & \(\times\)                   & \(\times\)                 & \(\times\)           & \(\times\)             & \(\times\)            & \(\times\)            & \(\times\)                 & {\small 722}                    & 4              \\ \hline
24          & \textit{\scriptsize versatile\_panel\_get\_modes}                    / \textit{\scriptsize tpg110\_get\_modes}                     & \(\times\)               & \(\times\)                   & 1                  & \(\checkmark\)           & \(\times\)             & \(\checkmark\)            & \(\times\)            & 31                 & 2                      & 2              \\ \hline
25          & \textit{\scriptsize ntb\_transport\_register\_client\_dev}           / \textit{\scriptsize locomo\_init\_one\_child}               & \(\times\)               & \(\times\)                   & \(\times\)                 & \(\times\)           & \(\times\)             & \(\times\)            & \(\times\)            & {\small 585}                & {\small 885}                    & 1              \\ \hline
26          & \textit{\scriptsize i3c\_master\_register\_new\_i3c\_devs}           / \textit{\scriptsize pci\_alloc\_child\_bus}                 & \(\times\)               & \(\times\)                   & \(\times\)                 & \(\times\)           & \(\times\)             & \(\times\)            & \(\times\)            & {\small 318}                & {\small 220}                    & 4              \\ \hline
27          & \textit{\scriptsize ch9getstatus}                                    / \textit{\scriptsize ast\_udc\_getstatus}                    & \(\times\)               & \(\times\)                   & \(\times\)                 & \(\times\)           & \(\times\)             & \(\times\)            & \(\times\)            & 8                  & 5                      & 4              \\ \hline
28          & \textit{\scriptsize fsl\_mc\_device\_add}                            / \textit{\scriptsize pci\_register\_host\_bridge}            & \(\times\)               & \(\times\)                   & \(\times\)                 & \(\times\)           & \(\times\)             & \(\times\)            & \(\times\)            & 12                 & 3                      & 8              \\ \hline
29          & \textit{\scriptsize amdgpu\_dm\_connector\_add\_common\_modes}       / \textit{\scriptsize versatile\_panel\_get\_modes}           & \(\times\)               & \(\times\)                   & \(\times\)                 & \(\times\)           & \(\times\)             & \(\times\)            & \(\times\)            & 92                 & {\small 175}                    & 8              \\ \hline
30          & \textit{\scriptsize ptp\_ocp\_device\_init}                          / \textit{\scriptsize tegra\_xusb\_port\_init}                & \(\times\)               & \(\times\)                   & \(\times\)                 & \(\times\)           & \(\times\)             & \(\times\)            & \(\times\)            & \(\times\)                 & {\small 123}                    & 10             \\ \hline
31          & \textit{\scriptsize ptp\_ocp\_device\_init}                          / \textit{\scriptsize srp\_add\_port}                         & \(\times\)               & \(\times\)                   & \(\times\)                 & \(\times\)           & \(\times\)             & \(\times\)            & \(\times\)            & \(\times\)                 & {\small 325}                    & {\small 11$^\star$}             \\ \hline
\multicolumn{12}{l}{%
    \parbox{\dimexpr\linewidth-2\tabcolsep}{% 自动计算可用宽度
        \footnotesize\raggedright%
        Numbers represent the rankings, and the symbol `\(\checkmark\)' indicates that the bug can be detected by a corresponding tool. Eleven bugs have been assigned CVEs, i.e., \#3, \#5, \#6, \#7, \#8, \#11, \#20, \#24, \#27, \#28 and \#29. 
        The case marked with `$^\star$' is ranked just beyond the top 10 (at 11th), but was noted due to its immediate adjacency to \#30 for the same seed function. 
    }%
} \\
\end{tabular}
\end{table}

\subsection{Detecting New Bugs}
\label{section:bug detection}

We first show the effectiveness of \sysname{} in detecting new bugs. 
Through the CVE list and the issue list in the Linux repository, 
we collect 55 seed functions with known bugs and attempt to detect bugs in Linux v6.4-rc2 with \sysname{}. The first author, as a skilled auditor, spends on average two minutes reviewing the top 10 candidates for each bug query. 

Among the 55 seeds, 28 led to the discovery of 31 new bugs that have been confirmed by the kernel developers and 11 of the bugs have been assigned CVEs. No other new bugs were detected by the tools for the other 27 seeds. 
The results are detailed in Table~\ref{tab:table_all_results}. 

The competitors explicitly expose worse performance compared to \sysname{} in detecting new bugs. 
ReDeBug and VUDDY fail to identify any bugs from the seed functions.
SourcererCC discovers five ($\frac{5}{31} = 16.1\%$). 
As a similarity-based matching method, Infercode can always compute a similarity score between a target and the seed, i.e., every function would have a ranking. We label those functions ranked beyond 100th with a `$\times$' symbol in Table~\ref{tab:table_all_results}, indicating that they are typically ignored in a manual audit. 
Even so, Infercode finds only seven (22.6\%), with six (19.4\%) among the top 10. 
NIL performs the best among the first seven detectors listed in Table~\ref{tab:table_all_results}, with 11 (35.5\%) bugs detected. However, NIL got stuck on the large code base. To evaluate its capabilities, we created a subset containing the detected buggy functions and tested whether NIL could match them to the seed functions. 
FIRE, the relevant SOTA tool for bug detection based on sliced taint paths, finds only eight bugs, accounting for 25.8\% of the total. 
SICode requires manual seed slicing but fails to specify detailed rules. Consequently, we build the seed graph using the query slice obtained by \sysname{}, which may substantially affect the embeddings and subgraph isomorphism decisions. 
It only detects three bugs (9.7\%), two of which are also easily identifiable by other tools. 
The two UniXcoder models can find more bugs within the top 10 than the other seven comparative tools, 15 (48.4\%) and 19 (61.3\%), respectively. UniXcoder-POJ, fine-tuned on the C-based code clone dataset, performs better than the original model. 
% \semihide{Though}
We also observe that, though UniXcoder-POJ emits a lower ranking for some buggy functions (e.g., \#5 and \#8), UniXcoder fails to find three within the top 1000 in the new bugs (i.e., \#23, \#30 and \#31). 

From the \sysname{} column, 27 buggy functions (87\%) are ranked within the top 5, making them easy to audit and the bugs easy to spot. 
Three are ranked between eighth (\#28 and \#29) and tenth (\#30), in an acceptable scope that usually does not irritate the auditors. 
The last one (\#31) is a notable exception. Its ranking is lower than the predefined top 10 (Section~\ref{section:embedding ranking}), but it immediately follows \#30 for the same seed function and we noticed it. 
Another thing worth noting is that, when using the same seed function as in \#30, \sysname{} actually detected another real bug ranked first in Linux v6.4-rc2. 
However, when we reported it to the developers, we got a reply saying the same bug had been fixed in v6.13-rc1 by someone else. 
Therefore, we do not include this bug in Table~\ref{tab:table_all_results}. 
Comparing the last two columns, we observe that, when the function-level similarity makes some buggy candidate functions neglected, \sysname{} significantly promotes the probability of the involved bugs being spotted, e.g., \#5, \#8, \#23 and \#25. 
In fact, compared to UniXCoder-POJ, \sysname{} improves the ranking for 28 buggy functions (90.3\%), demonstrating the effectiveness of the feature slice-based similarity measurement.

\begin{table}[t]
\centering
\caption{Bug detection results on the benchmark with a strict top-10 candidate audit for \sysname{}. }
\label{tab:table_all_results_76pairs_20250829}
\setlength{\tabcolsep}{4.8pt}
\begin{tabular}{|c|c|c|c|c|c|c|c|c|c|c|c|}
%\begin{tabularx}{\textwidth}{|c|c|c|c|c|c|c|c|c|c|c|c|}
\hline \textbf{Datasets} & 
    \rotatebox{90}{\textbf{\# Pairs}} & 
    \rotatebox{90}{\textbf{ReDeBug}} & 
    \rotatebox{90}{\textbf{SourcererCC}} & 
    \rotatebox{90}{\textbf{Infercode}} & 
    \rotatebox{90}{\textbf{NIL}} & 
    \rotatebox{90}{\textbf{VUDDY}} & 
    \rotatebox{90}{\textbf{FIRE}} & 
    \rotatebox{90}{\textbf{SICode}} & 
    \rotatebox{90}{\textbf{UniXcoder}} & 
    \rotatebox{90}{\textbf{UniXcoder-POJ}} & 
    \rotatebox{90}{\textbf{MATUS}} \\ \hline
% \textbf{New Bugs}
\textbf{D1}                  & 31           & 0\%           & 14.3\%                  & 19.4\%                & 35.5\%        & 0\%            & 22.9\%           & 9.7\%             & 48.4\%             & 61.3\%                 & 96.8\%         \\ \hline
% \textbf{Manually Collection}
\textbf{D2}           & 31              & 0\%              & 0\%                  & 19.4\%                & 19.4\%       & 9.7\%          & 12.9\%        & 9.7\%             & 38.7\%             & 35.5\%                 & 80.7\%         \\ \hline
%\textbf{From SICode}
\textbf{D3}                  & 14           & 14.3\%           & 0\%                  & 0\%                & 7.1\%        & 0\%            & 0\%           & 0\%             & 21.4\%             & 35.7\%                 & 35.7\%         \\ \hline
% \textbf{Overall (Excluding New Bugs)}
\textbf{D2+D3} & 45            & 4.4\%            & 0\%                  & 6.7\%                & 15.6\%       & 6.7\%          & 8.9\%         & 6.7\%             & 33.3\%             & 35.6\%                 & 64.4\%         \\ \hline
% \textbf{Overall (Including New Bugs)}
\textbf{D1+D2+D3} & 76            & 2.6\%            & 6.6\%                & 11.8\%                & 23.7\%       & 3.9\%          & 15.8\%        & 7.9\%             & 39.4\%             & 46.1\%                 & 77.6\%         \\ \hline

\end{tabular}
%\end{tabularx}
\end{table}

\subsection{Benchmark Performance}
\label{section:Detection Performance Comparison}

Since the new bugs are a subset of the benchmark, in this section, we focus on the performance on the benchmark. The results are shown in Table~\ref{tab:table_all_results_76pairs_20250829}. Be aware of that, to ensure a fair comparison, we strictly audit the top 10 candidates for each query in ranking-based methods such as \sysname{}. 

\sysname{} discovers 59 ($\frac{59}{76} = 77.6\%$) over the benchmark (D1+D2+D3) and exceeds all other tools on each subset. 
Even without considering the new bugs, \sysname{} successfully detects 64.4\% of existing bugs (D2+D3). 
Among the five bugs detected in D3 by \sysname{}, four are from OpenSSL. 
The two UniXcoder models continue to outperform the other seven tools on each subset and the whole benchmark, and NIL achieves the third-best performance among the nine. 
SICode fails to report any bugs on D3, the dataset from its paper, for the same reason as in Section~\ref{section:bug detection}. 
In addition, four other tools (SourcererCC, Infercode, VUDDY and FIRE) emit no bugs at all on D3. 
However, the overall results (on D1+D2+D3) are consistent with those in Table~\ref{tab:table_all_results}. 

We manually inspect the 17 missed bugs by \sysname{}, especially the limited detection effectiveness on D3. Six are excluded as they fall outside the top-1000 candidates during the screening phase (five from D3); eight are affected by an excessive number of similar operations (four from D2 and three from D3), which dilute their ranking scores; one is due to parsing inaccuracies (from D3); one is missed because of a single slice statement; and one resulted from incorrect identification of kVars and rStmts in the seed function. 
While the other causes are easy to understand, an excessive number of similar operations can yield lots of similar slices, and a minor deviation in the slices (e.g., variable or function names) may affect the embeddings. As a result, the buggy slices may get slightly lower similarity with the query and fall outside the audit scope. 

On the benchmark, only three bugs are detected by the competitors but missed by \sysname{}. VUDDY detects the last case among the 17 (failing kVar/rStmt identification). UniXcoder-POJ detects the one with parsing issues and the other one with lower ranking by \sysname{} (18th) due to similar operations. 

Overall, the benchmark experiment has well demonstrated the effectiveness of \sysname{} in bug detection.

\subsection{Efficiency}
\label{section:Efficiency}

Table~\ref{tab:table_all_timeConsumingCompare} presents a comparison of the average time cost per query across different tools on Linux v6.4-rc2. 
Generally, we count the offline preprocessing and online matching separately, if applicable. 
Though the offline step may cost a lot, online matching is often fast. 
For example, Infercode, SICode and UniXcoder(-POJ) take hours to embed the functions offline but spend only a few minutes to finish a query on $\thicksim$300k targets. 
VUDDY exhibits faster online matching performance, averaging 15 seconds per query. 
NIL, however, fails to process the whole Linux code base. Hence, we only evaluate its speed on the small subset mentioned in Section~\ref{section:Detection Performance Comparison}, which is about 8 seconds per query. 

% \usepackage{multirow}
\begin{table}[t]
\centering
\caption{%Time consumption
Time consumption. Online cost is measured per query on average. ScrCC is short for SourcererCC and UniXcoder-POJ has the same time cost as UniXcoder. 
}
% \setlength{\extrarowheight}{3pt} % 增加行距3pt
%\small
%\scriptsize
%\tiny
% \def\oldtabcolsep\tabcolsep
% \setlength{\tabcolsep}{4pt}
\setlength{\tabcolsep}{4.6pt}
% \begin{tabular}{c}
% \begin{minipage}{\linewidth}
\begin{tabular}{|c|c|c|c|c|c|c|c|c|}
\hline
{}                  & {\textbf{ReDeBug}}               & {\textbf{ScrCC}}   & {\textbf{Infercode}} & {\textbf{NIL}}                & {\textbf{VUDDY}} & {\textbf{FIRE}}         & \textbf{SICode}        & \textbf{UniXcoder} \\ \hline
{\textbf{Offline}}  & {\multirow{2}{*}{19m44s}}        & {\multirow{2}{*}{5m33s}} & {4h6m54s}            & {\multirow{2}{*}{\makecell{N/A \\(8s)}}} & {2h50m45s}       & {\multirow{2}{*}{\makecell{58m\\34s}}}       & {5h42m}  & 1h10m25s                 \\ \cline{1-1} \cline{4-4} \cline{6-6} \cline{8-9} 
{\textbf{Online}}   & {}       & {}  & {1m}                 & {}    & {15s}            & {}     & {4m39s}  & 1m    \\ \hline
\end{tabular}
% \end{minipage}

% \setlength{\tabcolsep}{7pt}
\setlength{\tabcolsep}{8pt}
% \begin{minipage}{\linewidth}
\begin{tabular}{|c|c|c|c|c|c|}
\hline
\multicolumn{6}{|c|}{\textbf{\sysname{}}}  \\ \hline
{\textbf{Offline}}  & {\textbf{\makecell{Candidate \\Screening}}} & {\textbf{\makecell{Pinpointing \\rStmt \& kVar}}}      & {\textbf{\makecell{Obtaining \\Feature Slices}}}        & {\textbf{\makecell{Matching \\ \& Ranking}}} & \textbf{\makecell{Total \\(Online)}}   \\ \hline
{42h1m32s} & {1m}     & {2m15s}  & {42s}      & {15s}   & 4m12s   \\ \hline
\end{tabular}
% \end{minipage}

% \end{tabular}
% \begin{tablenotes}
%     \footnotesize
%         \item Online cost is measured per query on average. 
% \end{tablenotes}
\label{tab:table_all_timeConsumingCompare}
\end{table}

As mentioned in Section~\ref{section:exp setup}, \sysname{} processes some steps offline, which takes about 42 hours, as shown in Table~\ref{tab:table_all_timeConsumingCompare}. 
As a one-time effort, it is acceptable and can be further accelerated by introducing parallel processing. 
For the online stages, though embedding vector-based similarity computation can be significantly sped up with GPU-enhanced batch processing, to ensure a fair comparison, we do the computation one by one. 
Given a seed function pair, \sysname{} takes negligible time to automatically identify the kVars and rStmts from the buggy function, so we omit the corresponding time cost in Table~\ref{tab:table_all_timeConsumingCompare}. 
The most time-consuming step is to pinpoint rStmts and kVars from the candidate functions, and the total online matching requires about four minutes. 
We deem it acceptable for a large code base such as Linux in practice, especially taking the detection performance in Table~\ref{tab:table_all_results} into account.

\subsection{Ablation Study}
\label{section:Ablation}

In this section, we conduct two groups of ablation study to evaluate the effectiveness of different model selection and different technical choices in \sysname{}. 

\subsubsection{Selecting Different Embedding Models}
\label{section:ablation:models}

We first evaluate how the fine-tuned model could affect the performance of \sysname{} compared to the original one. \sysname{} involves the embedding model at three steps, i.e., funciton-level screening, kVar/rStmt pinpointing and slice embedding. 
Applying either model (UniXcoder or UniXcoder-POJ) to the steps emits eight combinations in total. 
We present the combinations and the results on the benchmark in Table~\ref{tab:table_of_ablation-20250830}.

Clearly, UniXcoder-POJ has positive impact on each step. For example, with UniXcoder-POJ on screening and pinpointing (C5), the recall@10 achieves 75.0\%, higher than imposing it on screening only (C2: 67.1\%) or pinpointing only (C3: 68.4\%). 
The worst recall occurs when using UniXcoder for all the steps (C1), with only 64.5\% ($\frac{49}{76}$) bugs identified and 10 fewer than using UniXcoder-POJ for all steps (C8). 
Among the three steps, UniXcoder-POJ has the least impact on screening (2.6\% $\thicksim$ 6.6\%) and the most impact on pinpointing (3.9\% $\thicksim$ 7.9\%).
The result well demonstrates the effectiveness of the fine-tuned model as the embedding model.

\begin{table*}[t]
\centering
\caption{Ablation study result of selecting different embedding models on the benchmark. 
}
\setlength{\tabcolsep}{7.5pt}
\begin{tabularx}{\textwidth}{|c|c|c|c|c||c|c|c|c|c|}
\hline
\textbf{ID} & \textbf{SCR} & \textbf{PIN} & \textbf{EMB} & \textbf{Recall@10} & 
\textbf{ID} & \textbf{SCR} & \textbf{PIN} & \textbf{EMB} & \textbf{Recall@10}  
\\ \hline

C1           & $\circ$           &  $\circ$           & $\circ$           & 64.5\%                            &
C5          & $\bullet$           &  $\bullet$           & $\circ$           & 75.0\%                         \\ \hline
C2           & $\bullet$           &  $\circ$           & $\circ$           & 67.1\%                          &
C6          & $\bullet$           &  $\circ$           & $\bullet$           & 69.7\%                         \\ \hline
C3           & $\circ$           &  $\bullet$           & $\circ$           & 68.4\%                          &
C7          & $\circ$           &  $\bullet$           & $\bullet$           & 75.0\%                         \\ \hline
C4          & $\circ$           &  $\circ$           & $\bullet$           & 68.4\%                           &
C8          & $\bullet$           &  $\bullet$           & $\bullet$           & 77.6\%                       \\ \hline

% 新尝试的加脚注的： mk 20250530
\multicolumn{10}{l}{%
    \parbox{\dimexpr\linewidth-2\tabcolsep}{% 自动计算可用宽度
        \footnotesize\raggedright%
        Symbol $\bullet$ ($\circ$) indicates that UniXcoder-POJ (UniXcoder) is used for function-level screening (SCR), kVar/rStmt pinpointing (PIN) or feature-slice embedding (EMB).  
        Recall@10 indicates how many bugs are discovered within the top 10 candidates. 
    }%
} \\
%\end{tabular}
\end{tabularx}

\label{tab:table_of_ablation-20250830}
\end{table*}

\subsubsection{Taking Different Technical Choices}
\label{section:ablation:choices}

\begin{table}[t]
\centering
\caption{Ablation study result of taking different technical choices.}
\label{tab:table_of_ablation_2-20250830}
\begin{tabular}{|l|c|}
\hline
\textbf{Technical Choice} & \textbf{Recall@10} \\
\hline
Default \textsc{\sysname{}} & 77.6\% \\ \hline
\hspace{1em}(1) Preserving all changed as seed kVars/rStmts & 60.5\% \\ \hline
\hspace{1em}(2.1) Strict one step slicing & 68.4\% \\ \hline
\hspace{1em}(2.2) Unconstrained slice depth & 51.3\% \\ \hline
\hspace{1em}(3) Pinpointing candidate kVar/rStmt with Equation~\ref{eq:concatTokens} & 55.3\% \\ \hline
\hspace{1em}(4) Obtaining target slice with direct \texttt{[MASK]} mapping & 52.6\% \\ \hline
\end{tabular}
\end{table}

Furthermore, we conduct an additional study to evaluate the impact of the key technical choices.
The choices involve four aspects: (1) seed kVar/rStmt selection, (2) slicing depth, (3) target kVar/rStmt identification and (4) target slice acquisition. 

An alternative to the first aspect is a strategy that we preserve all changed statements and involved variables in the seed function as the seed rStmts/kVars and take the same pinpointing/slicing steps as \sysname{} to obtain feature slices. 
The second aspect involves two alternative options when the customized slicing in \sysname{} goes forward one more step in certain cases (see Section~\ref{section:slicing}): 
(2.1) Strict one-step slicing discards an additional step in Algorithm~\ref{alg:slicing} and forces the slicing depth to be one; (2.2) Unconstrained slicing depth adopts classic slicing algorithms without limiting the depth. 
The options for the third one are Equation~\ref{eq:concatTokens} or Equation~\ref{eq:maskBased} for kVar/rStmt pinpointing, as described in Section~\ref{section:pinpointing}. \sysname{} takes the latter one by default. 
For the last aspect, 
an alternative approach is to pinpoint the target rStmts one by one by using each statement in the query slice as the seed rStmt (along with the pre-collected seed kVars) while prohibiting slicing on the target function. 
In other words, the pinpointed rStmts form the target slice. 
We list the alternatives in the first column in Table~\ref{tab:table_of_ablation_2-20250830}. 
Note that, the alternative techniques are independently evaluated, i.e., only the tested technique is adopted by \sysname{} with all the others kept unchanged. 

\begin{figure}[t]
  \centering
  \includegraphics[width=\linewidth,clip,]{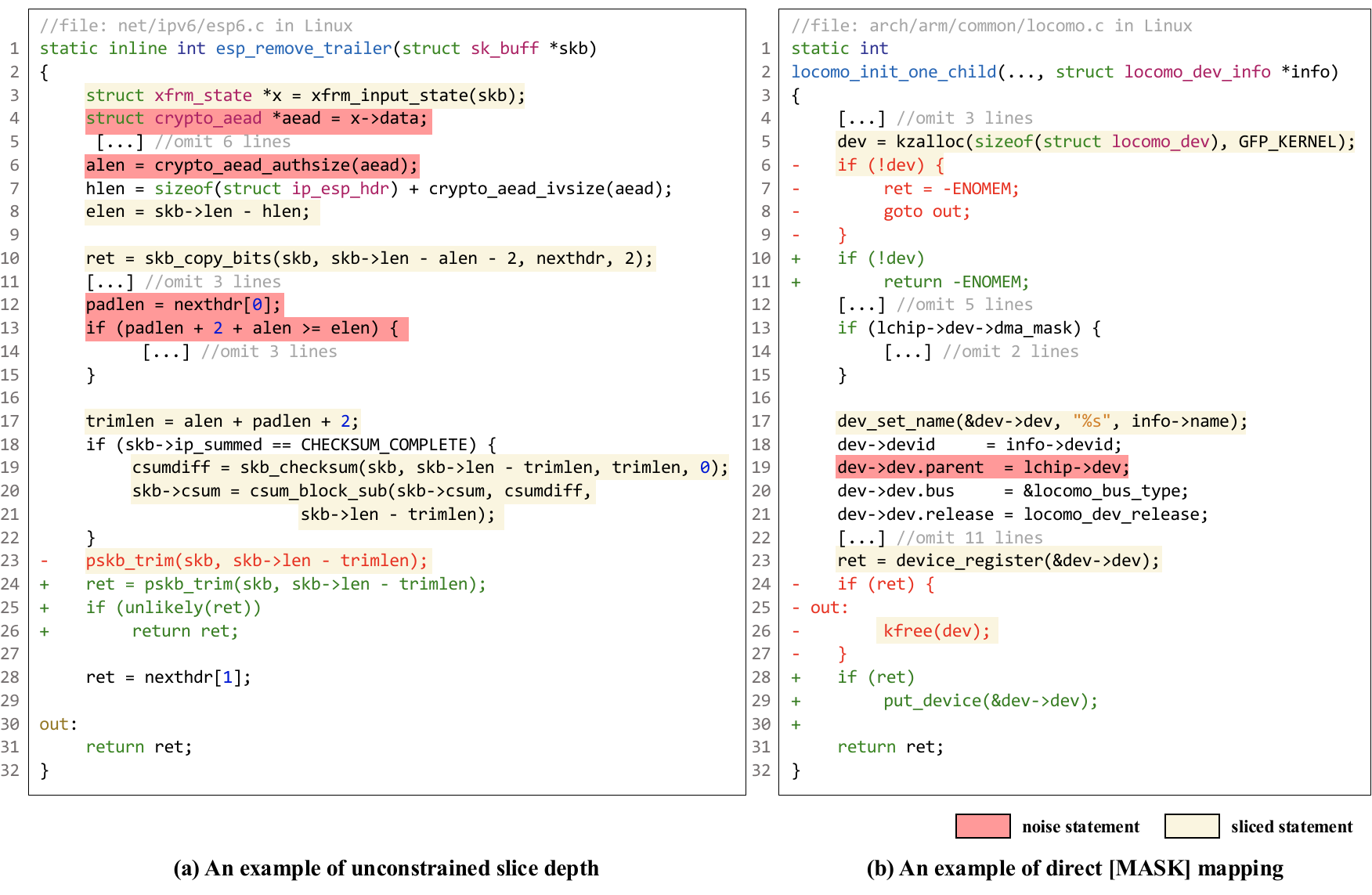}
  \caption{
  Examples of introducing noise statements into target slices with certain technique choices. 
  }
  \label{fig:ablation_examples_figures} 
\end{figure}

From Table~\ref{tab:table_of_ablation_2-20250830}, we can see that taking a different technical choice will reduces the recall, especially for the last three rows. One interesting observation is that, slicing less is better than slicing more, as extraneous noise statements will significantly influence the embedding-based slice similarity. Figure~\ref{fig:ablation_examples_figures}(a) presents a bug that is caught by \sysname{} (even with the strict one step slicing) but missed by adopting the (2.2) slicing strategy. 
It is also notable that, acquiring the target slice via direct \texttt{[MASK]} mapping (4) without further slicing can also introduce noise, causing declined recall (77.6\% $\rightarrow$ 52.6\%).  
Figure~\ref{fig:ablation_examples_figures}(b) shows an example, which is the same buggy function as  Figure~\ref{fig:motivating-example-1}(b). The direct mapping strategy introduces line 19, a noise statement of the bug, into the target slice, resulting in a degraded ranking of the target function. 

The study shows the necessity and effectiveness of the corresponding techniques \sysname{} adopts by default, which lead to better performance in similarity-based bug detection. 

\subsection{Parameter Sensitivity}
\label{section:Sensitivity}

We further examine how the number of candidate functions may affect the result in detecting new bugs in Linux. 
Besides the default value of 1000, three other conditions are evaluated, i.e., 500, 2000 and all. The last means that we do not perform the screening and leave all functions to subsequent steps. 
The result is shown in Table~\ref{tab:table_of_differentThreshold_comparison}. 
There is obviously a trade-off between efficiency and performance. 

The time cost is nearly linear with the number of candidate functions. Therefore, it takes over 14 hours on average to execute a query on the whole code base (the \textit{All} row), 207 times slower than the default that requires only 4 minutes. 

Keeping only the top 500 could miss four bugs, as expected from the UniXcoder-POJ column in Table~\ref{tab:table_all_results}. 
Analyzing the top 2000 emits the same result as the top 1000. Neither a bug in Table~\ref{tab:table_all_results} is missed, nor a new bug is spotted. 
For the last row, on the one hand, nine other bugs are uncovered in Linux v6.4-rc2, which have already been fixed in newer versions. 
They can also be detected by retaining the top 10000 as candidates, which, however, would take about 1 hour and 10 minutes per query. 
On the other hand, four bugs in Table~\ref{tab:table_all_results} drop their rankings below the top 15. We find that many similar but non-buggy functions are ranked higher than the buggy ones. 
Such a phenomenon will be further discussed in Section~\ref{section:Discussion}. 
Therefore, \sysname{}'s default choice balances efficiency and performance. 

% \usepackage{graphicx}
\begin{table}[t]
\centering
\caption{
Experiment result in detecting new bugs for different numbers of screened candidate functions. 
}
\begin{tabular}{|c|c|c|}
\hline
\textbf{\#Candidates} & \textbf{Time Cost} & \textbf{Results}            \\ \hline
Top 1000   & 4m12s        &  see Table~\ref{tab:table_all_results}   \\
\hline
Top 500   & 2m1s        &  \cancel{\#5}, \cancel{\#8}, \cancel{\#23} and \cancel{\#25}   \\ \hline
Top 2000   & 8m4s      & Same as Top 1000 \\ \hline
All    & 14h27m54s        & \cancel{\#11}, \cancel{\#27}, \cancel{\#30} and \cancel{\#31}; nine new that have been fixed.  \\
\hline
\end{tabular}
\label{tab:table_of_differentThreshold_comparison}
\end{table}

\section{Discussion}
\label{section:Discussion}

While \sysname{} has demonstrated its effectiveness in bug detection, there are some points that need further discussion.

\textbf{Automated or manual identification of seed kVar and rStmt.}
\sysname{} implements an automated identification of the seed kVars and rStmts, which enhances analysis efficiency and reduces the usage barrier of the tool. Human expertise can help \sysname{} handle the hard cases that may fail in an automated heuristic-based process, e.g., the one \sysname{} incorrectly recognized the seed kVar/rStmt and caused a missed bug (see Section~\ref{section:Detection Performance Comparison}). However, in our experiment, \sysname{} succeeds in identifying the seed kVars/rStmts in 98.7\% ($\frac{75}{76}$) cases, achieving nearly identical effectiveness to that of manual selection. 

\textbf{Limitations of the kVar/rStmt-based method.}
Though the experiments have well demonstrated the effectiveness of identifying kVar/rStmt in the seed and target functions, it has certain limitations in corner cases due to the code complexity. 
First, if the patch introduces entirely new statements and new variables that exhibit no interaction with any existing code, the current approach will fail to spot any seed kVar/rStmt for further analysis. 
Second, term frequency-based kVar identification may exclude the real object of interest, or at least include useless variables, leading to a redundant query slice and failing the matching. 
Third, if a patch of one function involves multiple (types of) bugs, it is highly possible to emit a large feature slice that cannot match any other target slices, when a small target slice possesses only one similar bug with the query. 
% We can leverage the human expertise to manually address the above issues. 
We can leverage human expertise to manually and precisely obtain code slices, thereby addressing the issues mentioned above. 
In particular, we can split the query slice in the third case into multiple small slices to enhance the performance. 

\textbf{Task-specific fine-tuning.} 
Pre-trained models are often fine-tuned specifically for downstream tasks. In this study, we only tune the model on a dataset targeting code clone detection. Neither is the dataset strongly related to the target of evaluation or bug detection, nor do we perform fine-grained tuning for masked embedding and feature slicing matching. Typically we should do that, but in practice, we have observed sufficiently high performance with the simply tuned model in combination with other techniques, making task-specific tuning unnecessary. However, due to the complexity of real-world code bases, there may be some scenarios where task-specific tuning must be taken to achieve acceptable performance. We will explore such cases in the future. 

\textbf{Side-effect of fixed/non-buggy targets.}
In large code bases such as Linux, there are many semantically similar functions, and many of them are non-buggy or have been fixed for the same bug in a given seed function. Their similarity with the seed function may significantly influence the result, when too many literally/semantically similar root statements can be found in the code base and the non-buggy code is considered more similar to the known buggy one. In such a case, the one with an unknown bug can have a low ranking and be ignored for auditing. For example, in Linux v6.4-rc2, there are 545 statements semantically similar to the rStmt calling \texttt{of\_clk\_add\_hw\_provider} in a known bug. A new bug ranked 31-\textit{st} by \sysname{} has the correctly pinpointed rStmt invoking \texttt{devm\_of\_clk\_add\_hw\_provider}. Among the 30 functions with higher ranking than the new buggy one, 94\% of them expose similar feature slices with the query but they are all non-buggy. 

A possible solution to excluding the similar but non-buggy candidates is to leverage the fixed version of the known bug, as done in~\cite{xiao2020mvp}. Higher similarity with the fixed one may probably indicate a non-buggy one. 
However, as Zhang et al.~\cite{zhang2022hunting} has illustrated, the assumption that fixed code is more similar to non-buggy code is not applicable in many cases for embedding-based methods. 
Nonetheless, we think it represents a chance to reduce the influence of fixed/non-buggy targets, though purposeful fine-tuning of the embedding models may be required. We leave it as a future exploration. 

\textbf{LLM-assisted candidate screening and slice extraction.} 
The powerful generative large language models (LLMs) can be an alternative to screen the candidate functions, pinpoint rStmts and kVars, and extract feature slices in an end-to-end manner. Ideally, we can feed LLM with the query slice of a known bug and a target function, asking LLM to extract a feature slice semantically similar to the query or to discard the function if no candidate slice can be found. 
We selected 5 hard samples (bug pairs with low similarity) and 5 easy (bug pairs with high similarity) in Table~\ref{tab:table_all_results}, and evaluated popular LLMs including ChatGPT~\cite{achiam2023gpt}, Deepseek-R1~\cite{liu2024deepseek} and LLaMA-7B~\cite{touvron2023llama}. 
For easy examples, the models may show satisfactory performance in pinpointing rStmts and kVars, and generate proper candidate feature slices within one test. 
We also find that the larger the LLM, the more likely it is to produce an answer as expected. 
However, inconsistency occurs with the same inputs, indicating stability issues. 
For hard examples, the models can barely pinpoint correct rStmts and kVars, or extract satisfied feature slices. 
With limited computational resources, fine-tuning these LLMs becomes impractical for us. We therefore leave it as an open problem.

\section{Related work}
\label{section:Related-work}

\subsection{Code Clone Detection}

In general, a pair of functions \{f1, f2\} is identified as clones if they are similar according to some definitions. In the current researches, a variety of methods have been developed for code clone detection. Depending on the data structures employed, these methods can generally be classified into categories such as text-based~\cite{roy2008nicad,su2016identifying}, token-based~\cite{sajnani2016sourcerercc,CCAligner,kamiya2021ccfinderx,li2006cpminer,ragkhitwetsagul2019siamese,CCLearner}, tree-based~\cite{wu2022detecting,xu2024dsfm,gao2019teccd,astnn,jiang2007deckard,yahya2023clcd,koschke2006clone}, graph-based~\cite{zou2020ccgraph,zhao2018deepsim,wu2020scdetector,shan2023gitor,CCSharp}, hybrid-method~\cite{li2024prism,saini2018oreo,wang2023ccstokener,vislavski2018licca,7582748,10.1145/3395363.3397362,wei2017supervised}, etc. NICAD~\cite{roy2008nicad} employs text normalization techniques to identify potential changes as simple text differences, making it effective for detecting near-miss intentional clones. 
Amain~\cite{wu2022detecting} uses Markov chain models to transform ASTs into state transition matrices, followed by feature extraction and machine learning classification to identify similar code fragments scalably. 
%%%
DSFM~\cite{xu2024dsfm} enhances functional code clone detection by incorporating deep subtree interactions, comparing subtrees from the abstract syntax trees of code snippets to introduce finer-grained semantic similarity. TECCD~\cite{gao2019teccd} utilizes a tree embedding technique with word2vec~\cite{church2017word2vec} to convert ASTs into vector representations, improving the efficiency of tree-based code clone detection and achieving high precision and recall for detecting type I, II, and III clones. 
%%%
CCGraph~\cite{zou2020ccgraph} is a PDG-based code clone detector that employs graph kernels and a two-stage filtering strategy with characteristic vector measurement, followed by an approximate graph matching algorithm using the Weisfeiler-Lehman graph kernel~\cite{shervashidze2011weisfeiler} to detect semantic clones. 
%%%
Prism~\cite{li2024prism} proposes a code clone detection method that leverages behavior semantics from multiple architecture assembly codes to enhance program understanding, utilizing an embedding technique and a multi-feature fusion strategy to create a more expressive representation of code.
 
\subsection{Large Models Used in Code-Related Tasks}

Currently, identifying semantic clone pairs with low textual similarity remains a significant challenge. With the advent of Artificial Intelligence (AI), researchers are seeking to address this issue through AI methods. In recent years, the rapid advancements in the field of large models~\cite{feng2020codebert,guo2020graphcodebert,ding2023concord,touvron2023llama,achiam2023gpt,taori2023alpaca,platzer2021vicuna}, with their increasing complexity and capabilities, have also brought new developments to the task of code clone detection.

CodeBERT~\cite{feng2020codebert} utilizes a Transformer architecture and is pre-trained on a replaced token detection task, enabling it to learn from both NL-PL pairs and unimodal data. While not directly mentioned for code clone detection, CodeBERT~\cite{feng2020codebert}'s capability to understand code semantics could enhance the detection of semantic clones. It achieves great performance in NL-PL tasks, suggesting potential applications in improving code clone detection accuracy. 
%%%
GraphCodeBERT~\cite{guo2020graphcodebert} achieves breakthroughs in code clone detection by leveraging semantic-level code structure and introducing structure-aware pre-training tasks. 
%%%
CONCORD~\cite{ding2023concord} employs a self-supervised pre-training strategy that incorporates code clones and their deviants to enhance the learning of general-purpose code representations, improving performance on downstream software engineering tasks such as clone and bug detection with reduced pre-training resource requirements. 
Additionally, large models such as LLaMA~\cite{touvron2023llama}, GPT4~\cite{achiam2023gpt}, Alpaca~\cite{taori2023alpaca}, and Vicuna~\cite{platzer2021vicuna}, which have more parameters, are also being explored. Researchers have attempted to apply these models to the task of code clone detection. 

\subsection{Bug Detection Based on Code Clone}

In the field of bug detection, bug detection~\cite{kim2017vuddy,xiao2020mvp,xu2021interpretation,bowman2020vgraph,eschweiler2016discovre,xu2020patch,10.1145/3433210.3437533} based on code clone is instrumental in preventing the propagation of software defects. 
VGRAPH~\cite{bowman2020vgraph} employs a graph-based approach to detect vulnerable code clones robustly by utilizing code property triplets, consisting of relationships from contextual, vulnerable, and patched code. 
BugGraph~\cite{10.1145/3433210.3437533} addresses source-binary code similarity detection by first identifying the compilation provenance of the target binary and compiling the source code accordingly. It then employs a graph triplet-loss network on attributed control flow graphs to rank code similarity. This method bridges the gap between source and binary code analysis.
MVP~\cite{xiao2020mvp} proposes a program slicing technique, which extracts and matches bug and patch signatures for detecting recurring bugs with reduced false positives and negatives. 
%%%
ISRD~\cite{xu2021interpretation} utilizes a multi-level birthmark model incorporating function, basic block, and instruction levels, along with Minimum Branch Path representation and intent search based on anchor recognition. 
DiscovRE~\cite{eschweiler2016discovre} employs control flow graphs and numeric feature pre-filters for rapid bug detection across binary architectures. 
%%%

\section{Conclusion}

In this paper, we present \sysname{}, an effective and scalable bug detection approach based on similar code matching. \sysname{} extracts the root statements and key variables from 
a snippet with a known bug, and then emits a query slice representing the semantic feature of the bug. 
The root statements and key variables are %further 
used to pinpoint their counterparts in target functions, which in turn guide the candidate feature slice extraction. 
The pinpointing leverages the power of an encoder model, based on which we propose a masked embedding-based approach to identification in an end-to-end way.
Feature slices are embedded with a fine-tuned pre-trained code model, and the similarity between each candidate and the query is computed. 
The top-ranked code slices and their corresponding functions are audited to confirm unknown bugs. 
\sysname{} successfully detected 31 previously unknown bugs from the Linux kernel and outperformed nine competitors. Experiments also demonstrate the effectiveness of the technical choices in \sysname{} and the acceptable efficiency for detecting bugs in a large code base.

\section*{Data Availability} % \section*避免被自动编号

The code and data necessary to reproduce the findings and analyses in this paper are available in an anonymized repository at \url{https://anonymous.4open.science/r/Yi3gA0k1aGz890}. To protect the double-blind review process, the authors have omitted any identifying information. The complete, non-anonymized replication package will be made publicly available upon publication of this paper. 

\begin{acks}

The authors would like to thank the anonymous reviewers for their valuable comments. 
The work is supported in part by 
Beijing Natural Science Foundation under grant 4262027, 
and National Natural Science Foundation of China (NSFC) under grants 62272465 and 62272464, 
and Public Computing Cloud, Renmin University of China. 

\end{acks}

\bibliographystyle{ACM-Reference-Format}
%\bibliography{sample-base}
\bibliography{references}

\appendix

% temporarily, mk 20250902
% \input{sections/appendix}

\end{document}